\documentclass[lettersize,journal]{IEEEtran}
\usepackage{amsmath,amsfonts}
\usepackage{algorithmic}
\usepackage{algorithm}
\usepackage{array}
\usepackage[caption=false,font=normalsize,labelfont=sf,textfont=sf]{subfig}
\usepackage{textcomp}
\usepackage{stfloats}
\usepackage{url}
\usepackage{verbatim}
\usepackage{graphicx}
\usepackage{bm}
\usepackage{amsthm,amsmath,amssymb}
\usepackage{mathrsfs}
\usepackage{soul,color,xcolor}
\usepackage{graphicx}
\usepackage{epstopdf}
\usepackage{cite}

\soulregister{\cite}7 
\def\BibTeX{{\rm B\kern-.05em{\sc i\kern-.025em b}\kern-.08em
    T\kern-.1667em\lower.7ex\hbox{E}\kern-.125emX}}
\usepackage{balance}
\begin{document}
\title{Double Self-Sustainable Reconfigurable Intelligent Surfaces Aided Wireless Communications}
\author{Ji Wang, Suhong Luo, Yixuan Li, Wenwu Xie, Xingwang Li, \IEEEmembership{Senior Member, IEEE}, \\
and Arumugam Nallanathan, \IEEEmembership{Fellow, IEEE}
	\thanks{Ji Wang, Suhong Luo and Yixuan Li are with the Department of Electronics and Information Engineering, College of Physical Science and Technology, Central China Normal University, Wuhan 430079, China (e-mail: jiwang@ccnu.edu.cn; lshhh@mails.ccnu.edu.cn; yixuanli@mails.ccnu.edu.cn).}
	\thanks{Wenwu Xie is with the School of Information Science and Engineering, Hunan Institute of Science and Technology, Yueyang 414006, China (e-mail: gavinxie@hnist.edu.cn).}
	\thanks{Xingwang Li is the School of Physics and Electronic Information Engineering, Henan Polytechnic University, Jiaozuo 454003, China (e-mail: lixingwang@hpu.edu.cn).}
	\thanks{Arumugam Nallanathan is with the School of Electronic Engineering and Computer Science, Queen Mary University of London, London E1 4NS, U.K. (e-mail: a.nallanathan@qmul.ac.uk).}}

\markboth{}%
{How to Use the IEEEtran \LaTeX \ Templates}

\maketitle

\begin{abstract}
A double self-sustainable reconfigurable intelligent surfaces (RISs) assisted multi-user multiple input multiple output (MIMO) system is investigated, where two RISs are equipped with energy harvesting circuit to achieve self-sustainable transmission. We aim to minimize the transmission power at the base station (BS), while guaranteeing the quality of service (QoS) requirements of the users and meeting the power consumption requirements of the RISs. In order to address the formulated coupled non-convex problem, we employ a block coordinate descent (BCD) algorithm based on the penalty-based method and successive convex approximation (SCA) to alternatively optimize the active beamforming at the BS and the phase shifts, as well as amplitude coefficients of two RISs. Simulation results show that the required power consumption at the BS for the proposed double self-sustainable RISs system is significantly reduced compared to conventional RIS systems.
\end{abstract}


\section{Introduction}
The sixth-generation (6G) communication system has garnered increasing attention due to its expansive coverage, broad spectrum reach, diverse applications, and enhanced security features \cite{8869705}. Anticipated to surpass current communication capacities offered by existing technologies. 6G communication poses challenges such as high hardware costs and significant energy consumption. To address these challenges, the exploration of more efficient methods to utilize spectrum resources is an inevitable trend for the future wireless communications. Fortunately, reconfigurable intelligent surface (RIS) has been recognized as a promising candidate solution  for empowering communication performance, owing to their advantages in easy deployment, high array gain, and low power consumption \cite{8981888,10347404}. Through the control of reflection coefficients on all reflection elements, RIS enables the creation of favorable energy and information transmission environments.

By utilizing the distinctive passive beamforming properties, extensive efforts have been dedicated to apply the RIS to various communication scenarios \cite{9086766}. In the conventional  RIS systems, only the single RIS is deployed to enhance communications, leading to limited coverage due to the passive nature of the RIS \cite{112}. Consequently, a cooperative double-RIS architecture has been explored to further improve the passive beamforming gain \cite{111,114}. A strategy involving deployment of two distributed RISs in proximity to the base station (BS) and a cluster of nearby users is examined. The authors in \cite{111} proved that double-RIS systems can achieve superior performance gains compared to single-RIS systems. Additionally, the authors in \cite{114} introduced double-active-RIS into wireless communication system to enhance the achievable rate.

Nevertheless, in the aforementioned works, RIS is assumed to be a purely passive surface \cite{111,114}. While these works suggest that the power consumption of each RIS element is minimal, the cumulative power consumption of the entire RIS becomes non-negligible, particularly when the amount of RIS elements is massive \cite{8741198}. Consequently, energy harvesting technology can be integrated into RIS to provide the required power of RIS operation, which is termed as a self-sustainable RIS architecture \cite{9214497}. Notably, the consideration of a double self-sustainable RISs-aided wireless communication system has not been explored in existing research.



Motivated by the above observations, we consider a double self-sustainable RISs aided multi-user multiple input multiple output (MIMO) communication system. We aim to minimize the transmission power at the BS by jointly optimizing the active beamforming at the BS, as well as the phase shifts and amplitude coefficients of the RISs. For the beamforming optimization at the BS, we employ the successive convex approximation (SCA) framework to address the quality of service (QoS) and power consumption constraints of the RISs. For the phase shifts of the RISs, we transform non-convex constraints through the penalty-based method. Subsequently, we propose a block coordinate descent (BCD) algorithm to solve the non-convex problem. Simulation results indicate that double self-sustainable RISs can achieve lower power consumption at the BS compared to conventional RIS systems.


\section{System Model And Problem Formulation}
\subsection{System Model}
In Fig.1, we consider a multi-user MIMO communication system aided by double self-sustainable RISs. The system comprises a BS with $M$ antennas, RIS1 deployed at the BS side with ${N_1}$ reflection elements, RIS2 deployed at the users side with ${N_2}$ reflection elements, and $K$ single-antenna users. To achieve self-sustainable transmission of the RISs, the RISs are installed with energy harvesting circuit to harvest energy from the BS. The channel from the BS to RIS1, the BS to RIS2, RIS1 to RIS2, RIS1 to the $k$-th user, and RIS2 to the $k$-th user are denoted as ${{\bf{H}}_1} \in {\mathbb{C}^{M \times {N_1}}}$,  ${{\bf{H}}_2} \in {\mathbb{C}^{M \times {N_2}}}$, ${\bf{D}} \in {\mathbb{C}^{{N_1} \times {N_2}}}$, ${{\bf{h}}_{1,k}} \in {\mathbb{C}^{{N_1} \times 1}}$, ${{\bf{h}}_{2,k}} \in {\mathbb{C}^{{N_2} \times 1}}$, while $k \in {\cal K} = \left\{ {1, \cdots ,K} \right\}$. Let ${{\bm{\theta }}_1} = {\left[ {{\theta _{1,1}}, \ldots ,{\theta _{1,{N_1}}}} \right]^T}$, and ${{\bm{\theta }}_2} = {\left[ {{\theta _{2,1}}, \ldots ,{\theta _{2,{N_2}}}} \right]^T}$ represent phase shifts of the reflection elements on RIS1 and RIS2, respectively. The transmitted signal from the BS is given by
\begin{flalign}
{\bf{x}} = \sum\limits_{k = 1}^K {{{\bf{w}}_k}{{s_k}}},
\end{flalign}
where ${{s_k}}$, ${{\bf{w}}_k} \in {\mathbb{C}^{M \times 1}}$ represent the baseband signal and the beamforming vector sent to the $k$-th user, respectively.
\begin{figure} [!t]
\centering
\includegraphics[width=0.35\textwidth]{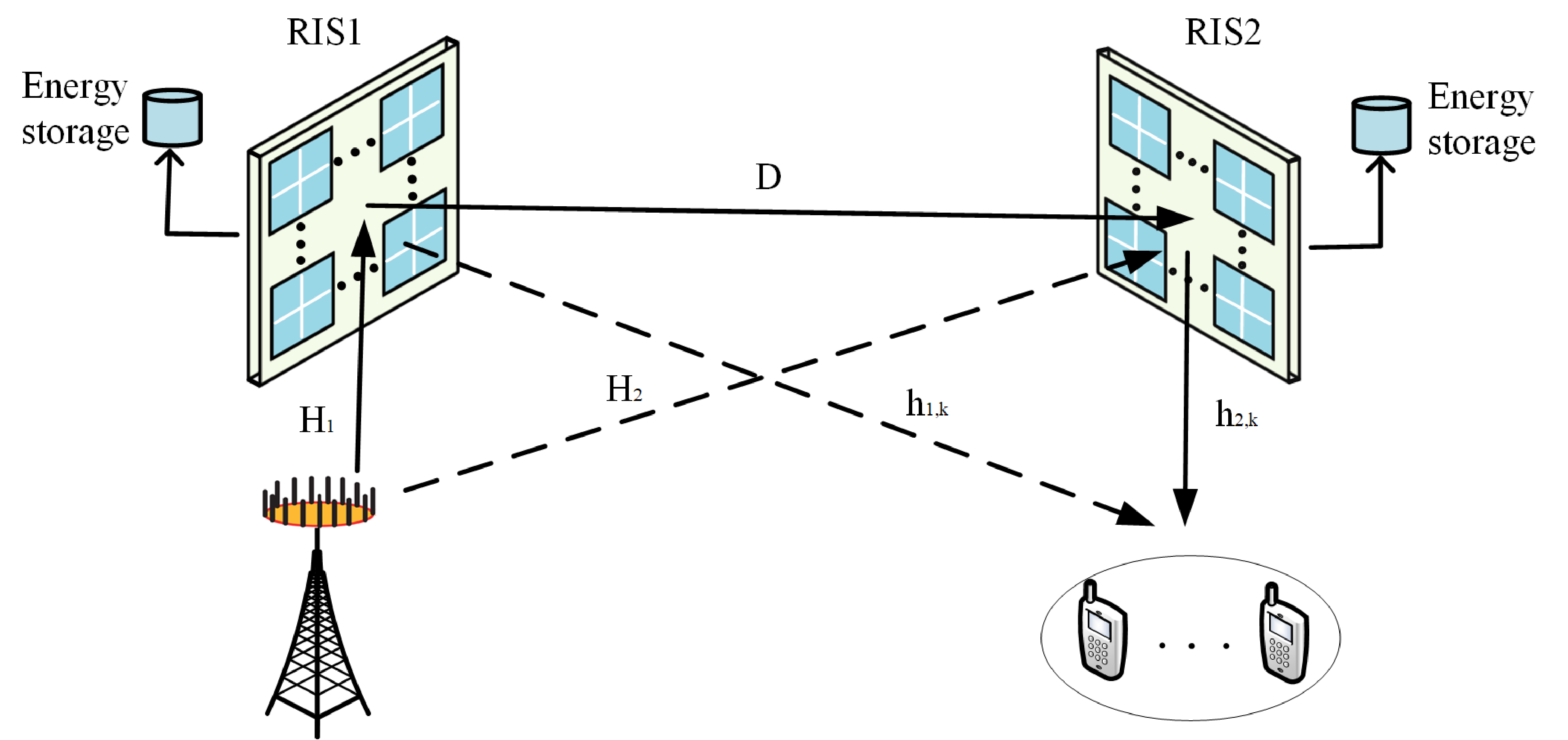}
\caption{A double self-sustainable RISs assisted multi-user MIMO system.}
\end{figure}

In this work, the RISs adopt power splitting (PS) scheme \cite{9214497}, where all elements of the RISs harvest a fraction of energy from the incident signal, and the remaining energy is reflected. Let ${\beta _1},{\beta _2}$ denote amplitude coefficients of the RISs. To reduce design complexity, all elements of the RISs are assumed to have the same amplitude coefficient. Thus, the received signal at the $k$-th user is expressed as
\begin{flalign}
{y_k} = {\bf{g}}_k^H{\bf{x}} + n,
\end{flalign}
\begin{flalign}
\!\!\!\!{{\bf{g}}_k} \!=\! {{\beta _1}\!{\bf{H}}_1}\!{{\bm{\Theta }}_1}\!{{\bf{h}}_{1,k}} \!+\! {{\beta _2}\!{\bf{H}}_2}{{\bf{\Theta }}_2}{{\bf{h}}_{2,k}} \!+\! {{\beta _1}{\bf{{\bf H}}}_1}{{\bf{\Theta }}_1}{\bf{D}}{{\bf{\Theta }}_2}{{\bf{h}}_{2,k}{\beta _2}},
\end{flalign}
which represents the joint channel from the BS to the $k$-th user and is composed of the double-hop channel and two single-hop channels. ${{\bm{\Theta }}_1} = {\rm diag}\left( {{{\bm{\theta }}_1}} \right){\mathbb{C}^{{N_1} \times {N_1}}}$, ${{\bf{\Theta }}_2} = {\rm diag}\left( {{{\bm{\theta }}_2}} \right){\mathbb{C}^{{N_2} \times {N_2}}}$ denote diagonal phase shift matrixes of RIS1 and RIS2, respectively. $n \sim {{\cal N}_c}\left( {0,{\sigma ^2}} \right)$ is the noise vector at the $k$-th user, where ${{\sigma ^2}}$ is the noise power \cite{112}. Let ${{{\bf{\tilde D}}}_k}{\bf{ = D}}{\rm diag}\left( {{{\bf{h}}_{2,k}}} \right) = \left[ {{{{\bf{\tilde d}}}_{1,k}}, \cdots ,{{{\bf{\tilde d}}}_{{N_2},k}}} \right]$. ${{\bf{R}}_{\lambda ,k}} = {{\bf{H}}_\lambda }{\rm diag}\left( {{{\bf{h}}_{\lambda ,k}}} \right)$ denotes the channel from the BS to ${\rm{RIS\lambda }}$ to the $k$-th user via ${\rm{RIS\lambda }}$, where $\lambda  \in \left\{ {1,2} \right\}$. The joint channel can be rewritten as
\begin{flalign}
{{\bf{g}}_k} &= {\beta _1}{{\bf{H}}_1}{{\bf{\Theta }}_1}{{{\bf{\tilde D}}}_k}{{\bm{\theta }}_2}{\beta _2} + \sum\limits_{\lambda  = 1}^2 {{\beta _\lambda }{{\bf{R}}_{\lambda ,k}}{{\bm{\theta }}_\lambda }}, \notag\\
 &= {\beta _1}{{\bf{H}}_1}\left[ {{\rm diag}\left( {{{{\bf{\tilde d}}}_{1,k}}} \right){{\bm{\theta }}_1}, \cdots ,{\rm diag}\left( {{{{\bf{\tilde d}}}_{{N_2},k}}} \right){{\bm{\theta }}_1}} \right]{{\bm{\theta }}_2}{\beta _2} \notag\\
 &~~~+ \sum\limits_{\lambda  = 1}^2 {{\beta _\lambda }{{\bf{R}}_{\lambda ,k}}{{\bm{\theta }}_\lambda }}, \notag\\
 &= \!{\beta _1}\!\!\sum\limits_{{n_2} = 1}^{{N_2}} \!{{{\bf{H}}_1}{\rm diag}\left( {{{{\bf{\tilde d}}}_{{n_2},k}}} \right){{\bm{\theta }}_1}} {\theta _{2,{n_2}}}{\beta _2} \!+\! \sum\limits_{\lambda  = 1}^2 {{\beta _\lambda }{{\bf{R}}_{\lambda ,k}}{{\bm{\theta }}_\lambda }}.
\end{flalign}

Let ${{\bf{Q}}_{n_2,k}} \!=\! {{\bm{{\rm H}}}_1}{\rm diag}(\!{{{\bf{\tilde d}}}_{n_2,k}}\!)$ represent the channel from the BS to the $k$-th user via RIS1 and RIS2. Both amplitude coefficients and the phase shifts are not considered in the single-hop and multi-hop channels mentioned above.
Accordingly, signal-to-interference-plus-noise ratio (SINR) of the received information at the $k$-th user  is formulated as
\begin{flalign}
{\Gamma _k} =& \frac{{{{\left| {{\bf{g}}_k^H{{\bf{w}}_k}} \right|}^2}}}{{\sum\limits_{i \ne k} {{{\left| {{\bf{g}}_k^H{{\bf{w}}_i}} \right|}^2} + {\sigma ^2}}}} = \frac{S}{{I + {\sigma ^2}}},
\end{flalign}
where $S$ denotes the received signal power, which is given by
\begin{flalign}
S \!\!=\!\! {\left|\! {\left( {{\beta _1}{{\bf{R}}_{1,k}}{{\bm{\theta }}_1} \!+\! {\beta _2}{{\bf{R}}_{2,k}}{{\bm{\theta }}_2} \!+\! \sum\limits_{n_2 = 1}^{{N_2}} {{{\bf{Q}}_{n_2,k}}{{\bm{\theta }}_1}{\beta _1}{\beta _2}{\theta _{2,n_2}}} } \right)\!{{\bf{w}}_k}} \right|^2\!}\!,
\end{flalign}
and $I$ denotes the interference power, which is given by
\begin{flalign}
I \!=\! \sum\limits_{i \ne k} {{{\left|\! {\!\left(\! {{\beta _1}{{\bf{R}}_{1,k}}{{\bm{\theta }}_1} \!+\! {\beta _2}{{\bf{R}}_{2,k}}{{\bm{\theta }}_2}\! +\! \sum\limits_{n_2 = 1}^{{N_2}} \!{{{\bf{Q}}_{n_2,k}}{\!{\bm{\theta }}_1}{\beta _1}{\beta _2}{\!\theta _{2,n_2}}} \!} \!\right)\!{{\bf{w}}_i}} \!\right|}^2}\!}\! .
\end{flalign}

Assuming the RISs adopt a linear energy harvesting model \cite{9214497}. Hence, harvested energy at RIS1 and RIS2 are given by
\begin{flalign}
{P_{\rm RIS1}} &= \eta \sum\limits_{k = 1}^K  \left( 1 - \beta _1^2 \right){{{\left\| {{{\bf{w}}_k}} \right\|}^2}}{\left\| {{{\bf{H}}_1}} \right\|^2},\\
{P_{\rm RIS2}} &= \eta \sum\limits_{k = 1}^K {{{\left( 1 - \beta _2^2 \right)\left\| {{{\bf{w}}_k}} \right\|}^2}} {\left\| {{{\bf{H}}_2}} \!\right\|^2} \notag\\
&+ \eta \!\sum\limits_{k = 1}^K \!{{{\beta _1^2\left( {1 - \beta _2^2} \right)\left\| {{{\bf{w}}_k}} \!\right\|}^2}}   {\left\| {{{\bf{H}}_{{1}}}{{\bf{\Theta }}_{{1}}}{\bf{D}}} \right\|^2},
\end{flalign}
where $\eta $ is the energy harvesting efficiency.

\subsection{Problem Formulation}
We aim to minimize the transmission power at the BS. The optimization is subject to QoS constraints for each user and energy consumption of the RISs. Thus, the problem is stated as follows
\begin{subequations}
\begin{eqnarray}
\!\!\!\!\left( {{\rm{P1}}} \right):\!\!\!\!&\mathop {\min }\limits_{\scriptstyle{{\bf{w}}_k},{{\bm{\Theta }}_1},\hfill\atop
\scriptstyle{{\bm{\Theta }}_2},{{\beta _1}},{{\beta _2}}\hfill}\!\!\! &\sum\limits_{k = 1}^K {{{\left\| {{{\bf{w}}_k}} \right\|}^2}} \label{p1-1}\\
&\rm s.t.&{\Gamma _k} \ge {\overline \Gamma_k},\label{p1-2}\\
&&{P_{\rm RIS1}} \ge {N_1}\mu, \label{p1-3}\\
&&{P_{\rm RIS2}} \ge {N_2}\mu, \label{p1-4}\\
&&\sum\limits_{k = 1}^K {{{\left\| {{{\bf{w}}_k}} \right\|}^2}}  \le {P_{\max }}, \label{p1-5}\\
&&\big| {{e^{j{\theta _{1,{n_{_1}}}}}}} \big| = 1,\forall {n_1} = 1, \cdots ,{N_1}, \label{p1-6}\\
&&\big| {{e^{j{\theta _{2,{n_{_2}}}}}}} \big| = 1,\forall {n_2} = 1, \cdots ,{N_2}, \label{p1-7}\\
&&0 \le {{\beta _1}} \le 1,\forall {n_1} = 1, \cdots ,{N_1}, \label{p1-8}\\
&&0 \le {{\beta _2}} \le 1,\forall {n_2} = 1, \cdots ,{N_2}, \label{p1-9}
\end{eqnarray}
\end{subequations}
where ${\overline \Gamma  _k}$, $\mu$, and ${P_{\max }}$ denote the desired QoS for each user, energy consumption of each RIS element, and transmission power budget at the BS, respectively. Constraint (\ref{p1-2}) is QoS constraints, which means that the SINR of each user should satisfy the SINR target. Constraints (\ref{p1-3}) and (\ref{p1-4}) are energy consumption constraints, which means that energy consumption of each RIS is no more than harvesting energy of each RIS. Constraints (\ref{p1-5})$-$(\ref{p1-9}) represent transmission power budget and the limits of amplitudes and the phase shifts at the RISs, respectively.

\section{Algorithm Design}
 All variables are coupled in the formulated problem (P1), rendering it a non-convex problem that cannot be solved immediately. To decouple and resolve this non-convex problem, a BCD algorithm is proposed to joint optimize the active beamforming at the BS, amplitude coefficients, as well as phase shifts of the RISs.

\emph{Proposition 1:} Under the proper operation of the RISs, we provide optimal amplitude coefficients,
\begin{flalign}
{\beta _1} &= \sqrt {1 - \frac{{{N_1}\mu }}{{\eta (\sum\limits_{k = 1}^K {\left\| {{{\bf{w}}_k}} \right\|_2^2)\left\| {{{\bf{H}}_1}} \right\|_2^2} }}} ,\label{beta1}\\
{\beta _2} &= \sqrt {1 - \frac{{{N_2}\mu }}{{\eta (\sum\limits_{k = 1}^K {\left\| {{{\bf{w}}_k}} \right\|_2^2} )\left\| {{{\bf{H}}_2}} \right\|_2^2 + P}}}.\label{beta2}
\end{flalign}
\emph{Proof:} Please refer to the Appendix A.

We can divide (P1) into two subproblems when optimal amplitude coefficients are obtained, each of which is a convex problem. Two subproblems are as follows: 1) given ${{\bf{\Theta }}_1},{{\bf{\Theta }}_2}$ to optimize ${\bf{w}}_k$; 2) given ${\bf{w}}_k$ to optimize ${{\bf{\Theta }}_1},{{\bf{\Theta }}_2}$.

\subsection{Optimizes ${\bf{w}}_k$ With Fixed ${{\bf{\Theta }}_1}$ and ${{\bf{\Theta }}_2}$}
Given ${{\bf{\Theta }}_1}$ and ${{\bf{\Theta }}_2}$, we can write the subproblem as
\begin{subequations}
\begin{eqnarray}
\left( {{\rm{P2}}} \right):&\mathop {\min }\limits_{{{\bf{w}}_k}} &\sum\limits_{k = 1}^K {\left\| {{{\bf{w}}_k}} \right\|_2^2} \\
&\rm s.t.&(\rm \ref{p1-2}),(\rm \ref{p1-3}),(\rm \ref{p1-4}),(\rm \ref{p1-5}),
\end{eqnarray}
\end{subequations}
problem (P2) can not be solved directly because constraints (\ref{p1-2}), (\ref{p1-3}), (\ref{p1-4}) are non-convex. In this section, we introduce a precoding algorithm utilizing SCA to approximate the problem of minimizing transmission power as a convex optimization problem. We phase shifting the constraint (\ref{p1-2}) and then perform a first-order Taylor expansion at the given point with $n$-th iteration ${\bf{w}}_k^{\left( n \right)}$, the constraint (\ref{p1-2}) is equivalent to
\begin{flalign}
2{\bf{w}}_k^{\left( n \right)H}\!{{\bf{g}}_k}{\bf{g}}_k^H{{\bf{w}}_k} \!- \!{\bf{w}}_k^{\left( n \right)H}{{\bf{g}}_k}{\bf{g}}_k^H{\bf{w}}_k^{\left( n \right)}\! \ge \!{{\bar \Gamma }_k\!}\left(\! {\sum\limits_{i \ne k} \!{{{\left| {{\bf{g}}_k^H{{\bf{w}}_i}} \right|}^2}\! +\! {\sigma ^2}} }\! \right)\!,\label{p2-5}
\end{flalign}
similarly, constraints (\ref{p1-3}) and (\ref{p1-4}) are equivalent to
\begin{eqnarray}
&&\!\!\!\!\!\!\sum\limits_{k = 1}^K {2{\bf{w}}_k^{\left( n \right)H}{{\bf{w}}_k} - {\bf{w}}_k^{\left( n \right)H}{\bf{w}}_k^{\left( n \right)}}\ge \frac{{{N_1}\mu }}{{\eta \left( {1 - {\beta _1^2}} \right){{\left\| {{{\bf{H}}_1}} \right\|}^2}}}\label{p2-6},\\
&&\!\!\!\!\!\!\sum\limits_{k = 1}^K {2{\bf{w}}_k^{\left( n \right)H}{{\bf{w}}_k}\! - \!{\bf{w}}_k^{\left( n \right)H}{\bf{w}}_k^{\left( n \right)}}\notag\\
&&\ge \frac{{{N_2}\mu }}{{\eta \left( {1\! -\! {\beta _2^2}} \right){{\left\| {{{\bf{H}}_2}} \right\|}^2}\! + \!\eta {\beta _1^2}\left( {1\! -\! {\beta _2^2}} \right){{\left\| {{{\bf{H}}_1}{{\bf{\Theta }}_1}{\bf{D}}} \right\|}^2}}}\label{p2-7},
\end{eqnarray}
then problem (P2) can be reformulated as
\begin{subequations}
\begin{eqnarray}
\left( {{\rm{P2.1}}} \right):&\mathop {\min }\limits_{{{\bf{w}}_k}} &\sum\limits_{k = 1}^K {\left\| {{{\bf{w}}_k}} \right\|_2^2}, \\
&\rm s.t.&(\ref{p2-5}), (\ref{p2-6}), (\ref{p2-7}), (\rm \ref{p1-5}),
\end{eqnarray}
\end{subequations}
which can be solved directly.
\begin{algorithm}[!t]
\caption{SCA algorithm for solving problem (P2.1).}
\label{alg1}
\begin{algorithmic}[1]
\STATE \small Initialization: Given ${\bf{w}}_k^{\left( 0 \right)}$, and set $n = 0$.
\STATE \small \textbf{repeat}
\STATE \small \hspace{0.3cm}Given ${\bf{w}}_k^{\left( n \right)}$, update ${\bf{w}}_k^{\left( {n + 1} \right)}$ with the solution of (P2.1).
\STATE \small \hspace{0.3cm}$n = n+1$.
\STATE \small \textbf{until} Convergence.
\end{algorithmic}
\label{alg1}
\end{algorithm}

\subsection{Optimizes ${{\bf{\Theta }}_1}$ and ${{\bf{\Theta }}_2}$ With Fixed ${\bf{w}}_k$}
Given ${\bf{w}}_k$, the problem is modeled as
\begin{subequations}
\begin{eqnarray}
\left( {{\rm{P3}}} \right):&\mathop  {\min }\limits_{{{\bm{\Theta }}_1},{{\bm{\Theta }}_2}} &\sum\limits_{k = 1}^K {{{\left\| {{{\bf{w}}_k}} \right\|}^2}} \\
&\rm s.t.&(\rm \ref{p1-2}),(\rm \ref{p1-6}),(\rm \ref{p1-7}),
\end{eqnarray}
\end{subequations}
 obviously, due to the coupling between ${{\bf{\Theta }}_1}$ and ${{\bf{\Theta }}_2}$, problem (P3) remains non-convex. We optimize ${{\bf{\Theta }}_1}$ when fixing ${{\bf{\Theta }}_2}$, the problem can be modeled as
\begin{subequations}
\begin{eqnarray}
\left( {{\rm{P3}}{\rm{.1}}} \right):&\mathop {\min }\limits_{{{\bm{\Theta }}_1}} &\sum\limits_{k = 1}^K {{{\left\| {{{\bf{w}}_k}} \right\|}^2}} \\
&\rm s.t.&\frac{{{{\left| {{\bm{\theta }}_1^H{{\bf{q}}_{k,k}} + {{\bar q}_{k,k}}} \right|}^2}}}{{\sum\limits_{i \ne k} {{{\left| {{\bm{\theta }}_1^H{{\bf{q}}_{k,i}} + {{\bar q}_{k,i}}} \right|}^2} + {\sigma ^2}} }} \ge {{\bar \Gamma }_k},\label{3-4}\\
&&(\rm \ref{p1-6}),\label{3-5}
\end{eqnarray}
\end{subequations}
where ${{\bf{q}}_{k,i}} = {\left( {\sum\nolimits_{{n_2} = 1}^{{N_2}} {{{\bf{Q}}_{n_2,k}}{\beta _1}{\beta _2}{\theta _{2,{n_2}}}}  + {{\bf{R}}_{1,k}}{\beta _1}} \right)^H}{{\bf{w}}_i}$, ${{{{\bar q}}}_{k,i}} = {\left( {{\beta _2}{{\bf{R}}_{2,k}}{{\bm{\theta }}_2}} \right)^H}{{\bf{w}}_i}$. Due to constraint (\ref{3-4}), problem (P3.1) remains non-convex. Constraint (\ref{3-4}) is equivalent to
\begin{flalign}
{\rm tr}\left( {{{\bf{B}}_{k,k}}{{\bf{\Psi }}_1}} \right) \ge {{\bar \Gamma }_k}\left( {\sum\limits_{i \ne k} {{\rm tr}\left( {{{\bf{B}}_{k,i}}{{\bf{\Psi }}_1}} \right)}  + {\sigma ^2}} \right),
\end{flalign}
where
\begin{flalign}
{{\bf{B}}_{k,i}} = \left[ {\begin{array}{*{20}{c}}
{{{\bf{q}}_{k,i}}{\bf{q}}_{k,i}^H}&{{{\bf{q}}_{k,i}}\bar q_{k,i}^H}\\
{{{\bar q}_{k,i}}{\bf{q}}_{k,i}^H}&{{{\bar q}_{k,i}}\bar q_{k,i}^H}
\end{array}} \right]\;\!,\!\;{{\bf{\Psi }}_1} \!= \!{{\bm{\tilde \theta }}_1}{\bm{\tilde \theta }}_1^H\;\!,\!\;{{\bm{\tilde \theta }}_1} = \left[ {\begin{array}{*{20}{c}}
{{{\bm{\theta }}_1}}\\
1
\end{array}} \!\right]\!,
\end{flalign}
with ${{\bf{\Psi }}_1}\succeq0$ and $\rm rank\left( {{{\bf{\Psi }}_1}} \right) = 1$, problem (P3.1) can transform into
\begin{subequations}
\begin{eqnarray}
\!\!\!\!\!\!\left( {\rm P3.2} \right)\!:\!\!\!\!&\mathop {\min }\limits_{{{{\bf{\Psi }}_1}}} \!\!\!\!&\sum\limits_{k = 1}^K {\left\| {{{\bf{w}}_k}} \right\|_2^2}\label{3-6} \\
&{\rm{s}}{\rm{.t}}{\rm{.}}\!\!\!\!&{\rm tr}\left( {{{\bf{B}}_{k,i}}{{\bf{\Psi }}_1}} \right) \!\!\ge\!\! {{\bar \Gamma }_k}\!\left(\! {\sum\limits_{i \ne k} {{\rm{tr}}\left( {{{\bf{B}}_{k,i}}{{\bf{\Psi }}_1}} \right)} \! +\! {\sigma ^2}} \!\! \right)\!\!,\label{3-7}\\
&&\left[ {{{\bf{\Psi }}_{1\,n,n}}} \right] = 1,\label{3-8}\\
&&{{\bf{\Psi }}_1}\succeq0,\label{3-9}\\
&&\rm rank\left( {{{\bf{\Psi }}_1}} \right) = 1\label{3-10}.
\end{eqnarray}
\end{subequations}

Previous researches have often employed Gaussian randomization to handle the rank-one constraint and devise approximate solutions when addressing similar problems \cite{9214497}. However, this approach does not yield satisfactory results in the alternating optimization algorithm. Hence, we adopt a penalty-based method as an alternative \cite{9957071}. To be specific, the objective function is rewritten as
\begin{flalign}
\mathop {\min }\limits_{{{\bf{\Psi }}_1}} \sum\limits_{k = 1}^K {{{\left\| {{{\bf{w}}_k}} \right\|}^2} + \tau \left( {\rm tr\left( {{{\bf{\Psi }}_1}} \right) - \left\| {{{\bf{\Psi }}_1}} \right\|} \right)}\label{3-11}.
\end{flalign}

To solve the concave function (\ref{3-11}), we rewrite (\ref{3-11}) by employing the first-order Taylor expansion \cite{9957071}
\begin{subequations}
\begin{eqnarray}
\left( {\rm P3.3} \right):&\mathop {\min }\limits_{{{\bf{\Psi }}_1}}& \sum\limits_{k = 1}^K {\left\| {{{\bf{w}}_k}} \right\|_2^2 + \tau z\left( {{{\bf{\Psi }}_1}} \right)}\\
&{\rm{s}}{\rm{.t}}{\rm{.}}&(\rm \ref{3-7}), (\rm \ref{3-8}), (\rm \ref{3-9}),
\end{eqnarray}
\end{subequations}
where
\begin{flalign}
z\left( {{{\bf{\Psi }}_1}} \right)\! =\! \left( {\rm tr\left( {{{\bf{\Psi }}_1}} \right)\! \!-\! \!\left\| {{{{\bf{\hat \Psi }}}_1}} \!\right\|\! \!-\! \!\rm tr\left[ {{{\bm{\varphi }}_{1\max }}{\bm{\varphi }}_{1\max }^H\left( \!{{{\bf{\Psi }}_1} \!-\! {{{\!\bf{\hat \Psi }}}_1}\!} \!\right)} \right]}\! \right),
\end{flalign}
in which $\tau $ is penalty factor. ${{{\bm{\varphi }}_{1\max }}}$ denotes the eigenvector associated with the maximum eigenvalue of a matrix ${{{{\bf{\hat \Psi }}}_1}}$, which can be solved directly.

Next, the solution methods for ${{\bf{\Theta }}_1}$ and ${{\bf{\Theta }}_2}$ are symmetrical. By optimizing ${{\bf{\Theta }}_2}$ in the same manner, we can rewrite problem (P3) as
\begin{subequations}
\begin{eqnarray}
\left( {{\rm{P3}}{\rm{.4}}} \right):&\mathop {\min }\limits_{{{\bf{\Theta }}_2}}& \sum\limits_{k = 1}^K {{{\left\| {{{\bf{w}}_k}} \right\|}^2}} \\
&\rm s.t.&\frac{{{{\left| {{\bm{\theta }}_2^H{{\bf{p}}_{k,k}} + {{\bar p}_{k,k}}} \right|}^2}}}{{\sum\limits_{i \ne k} {{{\left| {{\bm{\theta }}_2^H{{\bf{p}}_{k,i}} + {{\bar p}_{k,i}}} \right|}^2} + {\sigma ^2}} }} \ge {{\bar \Gamma }_k},\label{3-12}\\
&&(\rm \ref{p1-7}),\label{3-13}
\end{eqnarray}
\end{subequations}
where ${{\bf{p}}_{k,i}} = {\left( {\left[ {{{\bf{Q}}_{1,k}}{{\bm{\theta }}_1}, \cdots ,{{\bf{Q}}_{{N_2},k}}{{\bm{\theta }}_1}} \right]{\beta _1}{\beta _2} + {{\bf{R}}_{2,k}}{\beta _2}} \right)^H}{{\bf{w}}_i}$, ${{\bar p}_{k,i}} = {\left( {{\beta _1}{{\bf{R}}_{1,k}}{{\bm{\theta }}_1}} \right)^H}{{\bf{w}}_i}$. We can solve (P3.4) by following the procedure outlined in (P3.1) to (P3.3).
\begin{algorithm}[!t]
\caption{BCD algorithm based on SCA and the penalty-based method.}
\label{alg1}
\begin{algorithmic}[1]
\STATE \small Initialization: Given ${\bm{\theta }}_1^{\left( 0 \right)},{\bm{\theta }}_2^{\left( 0 \right)},{{{\bf{w}}_k}^{\left( 0 \right)}}$, and set $i = 0$.
\STATE \small \textbf{repeat: outer loop}
\STATE \small \hspace{0.3cm}Obtained $\beta _1$ and $\beta _2$ from (\ref{beta1}) and (\ref{beta2}) with given ${\bm{\theta }}_1^{\left( i \right)},{\bm{\theta }}_2^{\left( i \right)}$, and ${{{\bf{w}}_k}^{\left( i \right)}}$.
\STATE \small \hspace{0.3cm}Solved (P2.1) for given ${\bm{\theta }}_1^{\left( i \right)},{\bm{\theta }}_2^{\left( i \right)}$ via the SCA algorithm, obtaining ${{{\bf{w}}_k}^{\left( {i + 1} \right)}}$.
\STATE \small \hspace{0.3cm}\textbf{repeat: inner loop}
\STATE \small \hspace{0.3cm}\hspace{0.3cm}Solved (P3.3) for given ${\bm{\theta }}_2^{\left( i \right)},{{{\bf{w}}_k}^{\left( {i + 1} \right)}}$ via the penalty-based method, obtaining ${\bm{\theta }}_1^{\left( {i + 1} \right)}$.
\STATE \small \hspace{0.3cm}\hspace{0.3cm}Solved (P3.4) for given ${\bm{\theta }}_1^{\left( i+1 \right)},{{{\bf{w}}_k}^{\left( {i + 1} \right)}}$ via the penalty-based method, obtaining ${\bm{\theta }}_2^{\left( {i + 1} \right)}$.
\STATE \small\hspace{0.3cm}\textbf{until} Convergence.
\STATE \small \hspace{0.3cm}Update $i = i + 1$.
\STATE \small \textbf{until} The maximum number of iterations ${i_{\max }}$ has been reached or ${\rm O} = \min {\left\| {{\bf{w}}_k^{\left( {i + 1} \right)}} \right\|^2} - \min {\left\| {{\bf{w}}_k^{\left( i \right)}} \right\|^2} \le \varepsilon$, $\varepsilon  > 0$.
\end{algorithmic}
\end{algorithm}

\subsection{The Convergence and Computational Complexity}
Firstly, for problem (P1), we have obtained optimal solutions for two subproblems, and the objective function monotonically decreases as the number of iterations increases. Secondly, to ensure normal operation of the system, there exists a lower bound for transmission power.

Calculating the complexity of two subproblems separately. In the subproblem of optimizing ${\bf{w}}_k$ based on SCA, the complexity is ${\cal O}\left\{ \left( {{M^3}{K^{3.5}}} \right){S_1}\right\}$, where ${S_1}$ is the number of iterations. In the subproblem of optimizing ${\bf{\Theta }}$ based on alternating optimization, the complexity is ${\cal O}\left\{ \left( \sqrt {K + N} \left( {{M^3}{K^3}} \right) \right){S_2}\right\}$, where ${S_2}$ is the number of iterations. In conclusion, overall complexity of the algorithm is ${\cal O}\left\{ \left( {\left( {{M^3}{K^{3.5}}} \right){S_1} + \sqrt {K + N} \left( {{M^3}{K^3}} \right){S_2}} \right)Z\right\}$, where $Z$ is the number of iterations in the outer loop.

\begin{figure*}
\setlength{\abovecaptionskip}{0pt}
\setlength{\belowcaptionskip}{-16pt}
\centering
\begin{minipage}[t]{0.32\linewidth}
    \centerline{\includegraphics[width=2.5in]{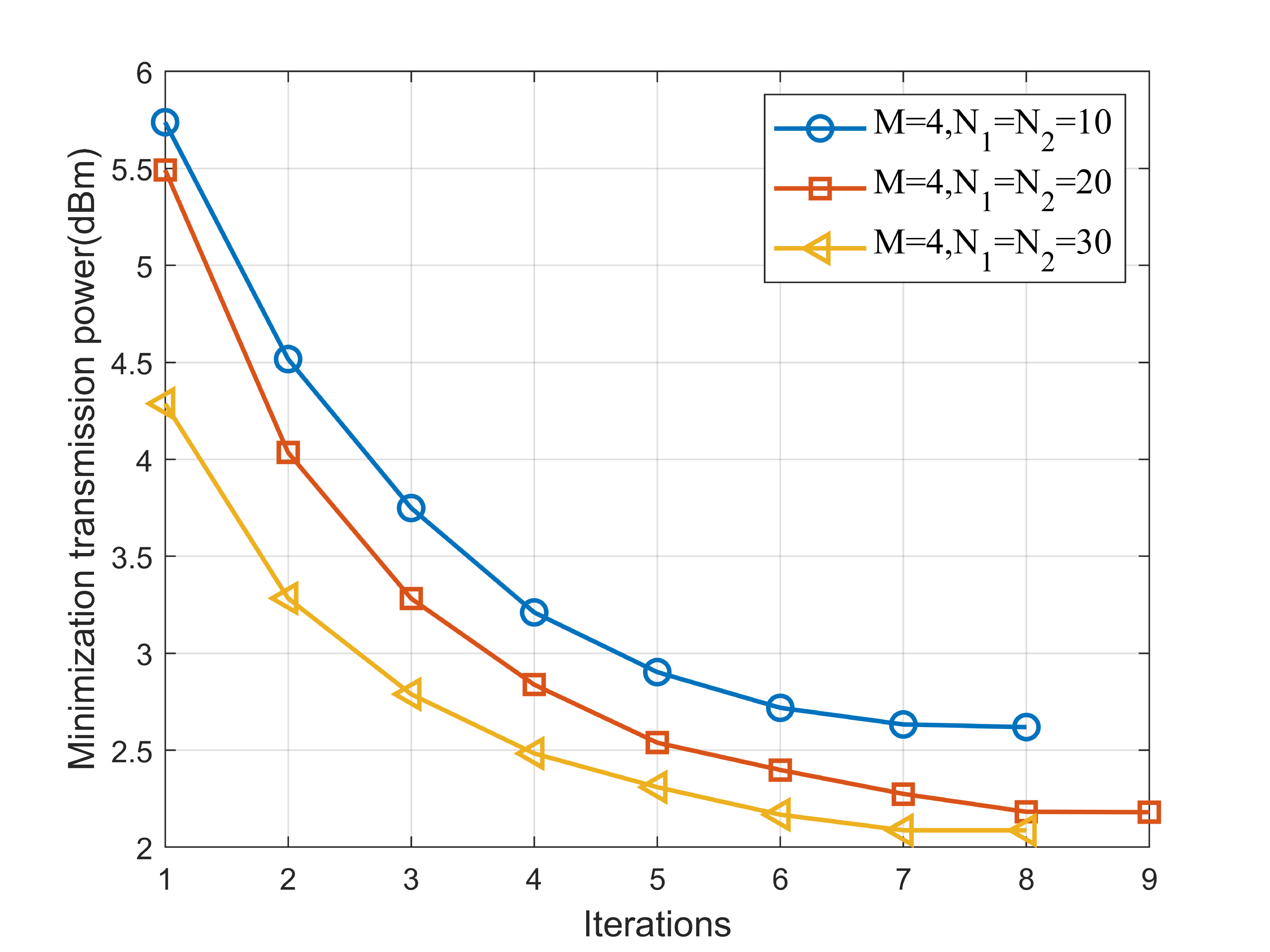}}
    \captionsetup{font={small}}
    \caption{Convergence of the proposed algorithm. }
    \label{fig2.}
\end{minipage}
\begin{minipage}[t]{0.33\linewidth}
    \centerline{\includegraphics[width=2.5in]{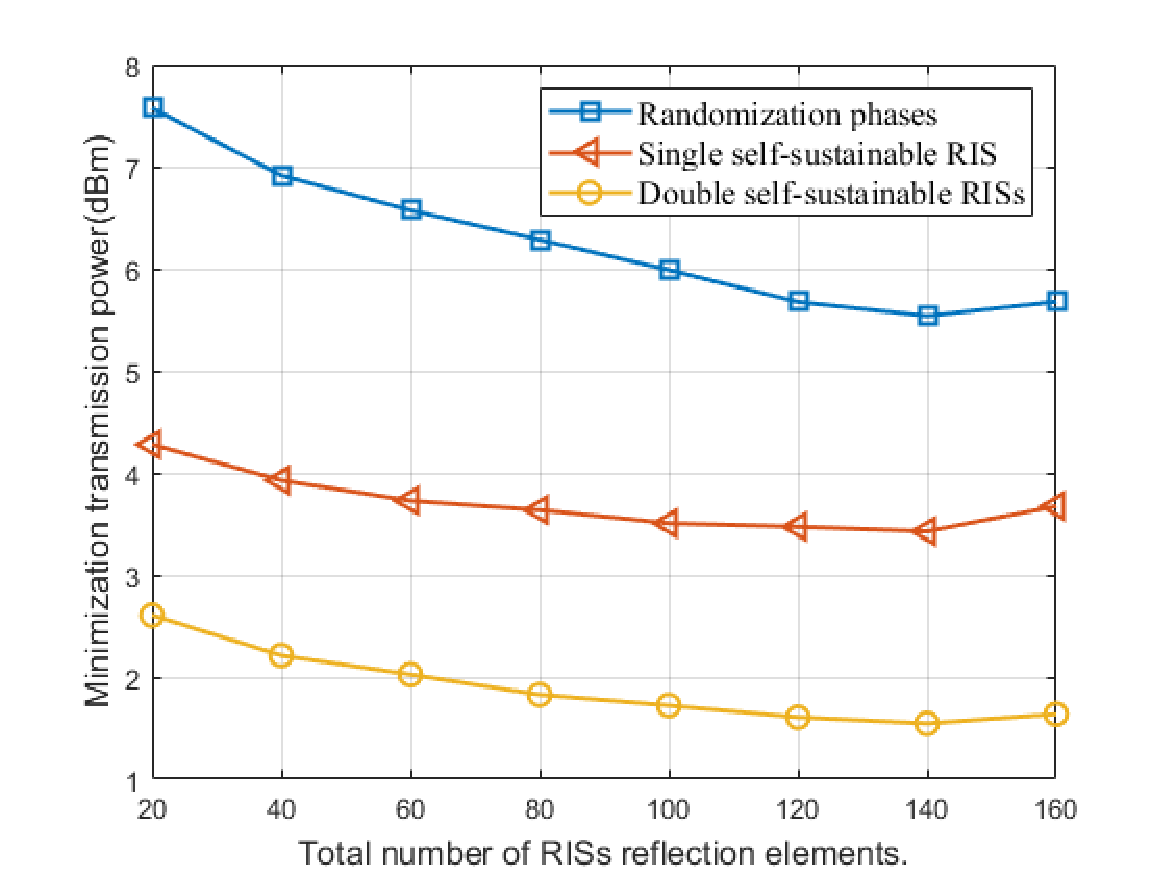}}
    \captionsetup{font={small}}
    \caption{Transmission power versus the total number of reflection elements with $N_1 = N_2$.}
    \label{fig3}
\end{minipage}
\begin{minipage}[t]{0.33\linewidth}
    \centerline{\includegraphics[width=2.5in]{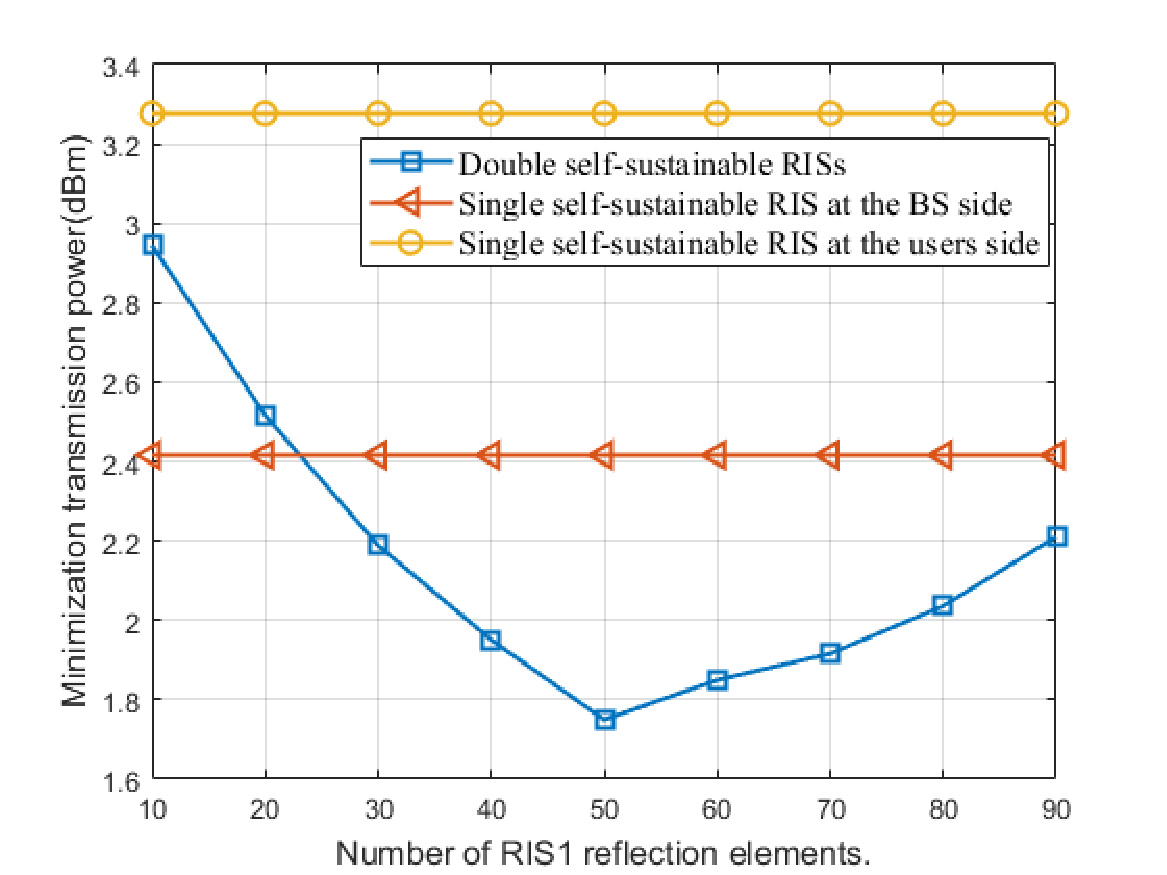}}
    \captionsetup{font={small}}
    \caption{Transmission power versus the reflection elements number of RIS1 with $N_1 + N_2 = 100$.}
    \label{fig4}
\end{minipage}
\end{figure*}
\section{Simulation Results}
We evaluate the performance of proposed system through simulation results. Assuming that the locations of the BS, RIS1 and RIS2 are given by $\left( {0,0} \right)$ m, $\left( {2,2} \right)$ m, $\left( {4,2} \right)$ m. The position of users are uniformly and randomly distributed at a cycle with 2 meters radius and centered at $\left( {6,0} \right)$ m \cite{9214497}. All channels in the considering system are Rician distribution \cite{10077727}, which can be given by
\begin{flalign}
{\bf{h}} \!= \!\sqrt {{\rho _0}{{\left( {\frac{d}{{{d_0}}}} \right)}^{ - \alpha }}} \left( {\sqrt {\frac{\kappa }{{\kappa  + 1}}} {{\overline {\bf{h}} }^{{\rm{LoS}}}} + \sqrt {\frac{1}{{\kappa  + 1}}} {{\overline {\bf{h}} }^{{\rm{NLoS}}}}}\! \right),
 \end{flalign}
   where $\rho_0 = -30$ dB is the large-scale fading coefficient with reference distance $d_0 = 1$ m. $d$ is the distance between two devices. $\kappa =5 $ is the Rician factor of the small-scale fading and $\alpha$ is the large-scale fading factor of the corresponding channel. The large-scale fading factor from the BS to RIS1 link, RIS1 to RIS2 link, and the BS to RIS2 link, RIS1 to the $k$-th user link, RIS2 to the $k$-th user link are set as $ {\alpha _{\rm BS,RIS1}} = 3.6$, ${\alpha _{{\rm RIS1},{\rm RIS2}}} = 2.2$, $ {\alpha _{\rm BS,\rm RIS2}} = 2.2$, ${\alpha _{{\rm RIS1},{\rm U_k}}} = 2.2$, ${\alpha _{{\rm RIS2},{\rm U_k}}}=2.2$, respectively. Unless otherwise stated, we set: $M = 4$, $K = 4$, $\eta  = 0.8$, ${{\bar \Gamma }_k} = 20\rm dBm$, $\mu  = 1\rm mW$, ${P_{\max }} = 40\rm dBm$, ${\sigma ^2}  =  - 110\rm dBm$, and $\tau  = 10$.

In Fig.2, we investigate the relationship between transmission power and the number of iterations. It can be observed that transmission power reaches a stable value after 8 iterations. In particular, with the number of reflection elements increases, transmission power decreases, because more reflection elements can achieve higher beamforming gains, thereby reducing the power consumption required by the BS.

In Fig. 3, we study the influence of the total number of reflection elements on the minimization transmission power. It is not difficult to notice that with the total number of reflection elements increases from 20 to 140, required transmission power decreases. This is due to a larger number of reflection elements can provide a higher beamforming gain. When the total number of reflection elements surpasses 140, the required transmission power increases. Because more reflection elements lead to higher power consumption at the RISs, which
will reduce the amplitude of reflected signals.

In Fig. 4, we analyze the effect of the number of reflection elements in RIS1 on the transmission power minimization with the aim of determining the optimal allocation strategy for elements in our considered system. We can find that the minimization transmission power required is minimum when $N_1 = 50$, which means that the minimization transmission power can be achieved by uniformly distributing the number of double RISs elements. For single-RIS scheme, the RIS deployed on the BS side performs better than the users side, because the RIS can harvest more energy closer to the BS.
\section{Conclusions}
In this work, we investigated double self-sustainable RISs assisted a MIMO communication system. Firstly, we formulated a transmission power minimization problem by jointly optimizing the precoded vector of the BS and the amplitudes and phase shifts of the RISs. Then, we designed a algorithm based on SCA to solve the QoS constraints and power consumption constraints. For solving phase shifts of the RISs, we employ the penalty-based algorithm to handle the rank-one constraint. Finally, simulation results indicated that comparing to the single RIS, deploying double RISs can obtain less minimization transmission power.

\appendices

\section{Proof of Proposition 1}
Firstly, to ensure that the RISs can assist with downlink communications, we must have ${P_{\rm RIS1}} \ge {N_1}\mu $, ${P_{\rm RIS2}} \ge {N_2}\mu $. The upper limit of  amplitude coefficients can be determined,
\begin{flalign}
{\beta _1} \le \sqrt {1 - \frac{{{N_1}\mu }}{{\eta (\sum\limits_{k = 1}^K {\left\| {{{\bf{w}}_k}} \right\|_2^2)\left\| {{{\bf{H}}_1}} \right\|_2^2} }}} ,\label{beta1}\\
{\beta _2} \le \sqrt {1 - \frac{{{N_2}\mu }}{{\eta (\sum\limits_{k = 1}^K {\left\| {{{\bf{w}}_k}} \right\|_2^2} )\left\| {{{\bf{H}}_2}} \right\|_2^2 + P}}},\label{beta2}
\end{flalign}
where $P = \eta (\sum\nolimits_{k = 1}^K {\left\| {{{\bf{w}}_k}} \right\|_2^2} ){\beta _1^2}{\left\| {{{\bf{H}}_{{1}}}{{\bf{\Theta }}_{{1}}}{\bf{D}}} \right\|^2}$ denotes the signal power received by RIS2 via RIS1. Secondly, to maximize the reflected power from the RISs, amplitude coefficients need to be set to their upper bounds at the optimal solution.
\bibliography{Reference_susRISs}
\bibliographystyle{IEEEtran}

\end{document}